\documentclass{article}

\makeatletter
\@namedef{ver@algorithmic.sty}{}
\makeatother

\usepackage{amsmath}
\usepackage{booktabs}
\usepackage{graphicx}
\usepackage{xcolor}
\usepackage{colortbl} % \cellcolor in tables
\usepackage{xspace}
\usepackage{tabularx}
\usepackage{array}
\usepackage{algorithm}
\usepackage{algpseudocode}
\algrenewcommand\algorithmicrequire{\textbf{Input:}}
\algrenewcommand\algorithmicensure{\textbf{Output:}}
\algrenewcommand\algorithmiccomment[1]{\hfill{\color{gray}$\triangleright$~#1}}
\newcommand{\concat}{\mathbin{\Vert}}
\usepackage{subcaption}
\usepackage{listings}
\usepackage[inline]{enumitem}
\lstdefinestyle{prompt}{%
  basicstyle=\ttfamily\scriptsize,
  columns=fullflexible,
  keepspaces=true,
  breaklines=false,
  showstringspaces=false,
  comment=[l]{\#},
  commentstyle=\normalfont\itshape\scriptsize,
  frame=none,
  aboveskip=2pt,
  belowskip=0pt,
  xleftmargin=0pt,
}
\lccode`\-=0 % hyphenate compound words only at their hyphens

\newcommand{\sysname}{\textsc{Mortis}\xspace}

\definecolor{exgrayF}{HTML}{E1E1E1}
\definecolor{exblueF}{HTML}{CCE0F5}
\definecolor{exorangeF}{HTML}{FCE0C3}
\definecolor{expurpleF}{HTML}{E6D6F0}
\newcommand{\para}[1]{\textbf{#1}\hspace{1em}\ignorespaces}
\usepackage{microtype}
\usepackage{hyperref}
\usepackage{xurl} % long reference URLs break instead of running into the margin

\usepackage[accepted]{mlsys2025}
\usepackage[varqu]{zi4}
\graphicspath{{figures/}{figures/evaluation/}}

\makeatletter
\renewcommand{\Notice@String}{Preprint.}
\renewcommand{\printAffiliationsAndNotice}[1]{%
\stepcounter{@affiliationcounter}%
{\let\thefootnote\relax\footnotetext{\hspace*{-\footnotesep}#1%
\forloop{@affilnum}{1}{\value{@affilnum} < \value{@affiliationcounter}}{ %
\textsuperscript{\arabic{@affilnum}}\csname @affilname\the@affilnum\endcsname}.
\ifdefined\mlsyscorrespondingauthor@text
Correspondence to: \mlsyscorrespondingauthor@text.
\fi
\ \\
\Notice@String
}}}
\makeatother

\mlsystitlerunning{\sysname}

\begin{document}

\twocolumn[
\mlsystitle{\sysname: Compiler-Guided KV-Cache Management for \\ LLM Agent Programs}

\begin{mlsysauthorlist}
\mlsysauthor{Xiaoyu Ma}{sjtu}
\mlsysauthor{Jayanaka L. Dantanarayana}{umich}
\mlsysauthor{Savini Kashmira}{umich}
\mlsysauthor{Baichuan Li}{umich}
\mlsysauthor{Lingjia Tang}{umich}
\mlsysauthor{Jason Mars}{umich}
\end{mlsysauthorlist}
\mlsysaffiliation{sjtu}{Shanghai Jiao Tong University}
\mlsysaffiliation{umich}{University of Michigan, Ann Arbor}
% Uncomment and fill in to print a "Correspondence to:" line in the footnote.
% \mlsyscorrespondingauthor{Xiaoyu Ma}{email@sjtu.edu.cn}

\vskip 0.3in

\begin{abstract}
Existing agent serving systems divide each prompt into a reusable prefix and an opaque suffix, managing the KV-cache at this coarse granularity without distinguishing the components within each part. We observe that prompt components differ in which LLM calls share their values and how long those values remain reusable, a property we term \textit{intra-prompt heterogeneity}. Our key insight is that, in frameworks that generate prompts from typed LLM function calls, these components correspond to program variables, allowing compiler analysis to recover their sharing patterns and lifetimes. We present \sysname, a compiler-serving co-design that combines this analysis with runtime observations to improve KV-cache management. \sysname reorders prompt components to expose additional prefix reuse and uses their predicted reuse and lifetimes to guide cache retention, eviction, and prefetching. Our evaluation shows that \sysname achieves up to 1.99$\times$ job completion time (JCT) speedup and 1.97$\times$ throughput improvement over SGLang, with geometric-mean speedups of 1.45$\times$ and 1.49$\times$, respectively.
\end{abstract}
]

\printAffiliationsAndNotice{}

% PDF metadata (set after \mlsystitle, which would otherwise override pdftitle).
\hypersetup{
  pdftitle={MORTIS: Compiler-Guided KV-Cache Management for LLM Agent Programs},
  pdfauthor={Xiaoyu Ma, Jayanaka L. Dantanarayana, Savini Kashmira, Baichuan Li, Lingjia Tang, Jason Mars},
  pdfsubject={LLM serving},
}
\mlsyskeywords{LLM serving, KV-cache management, LLM agents, LLM agent serving, compiler analysis, prefix caching}

\section{Introduction}
\label{sec:intro}

Agent applications solve tasks through sequences of dependent LLM calls that repeatedly consume common context, including instructions, input documents, and intermediate results. Yet repeated content does not automatically translate into reusable computation. Processing a prompt, known as \emph{prefill}, produces attention states stored in a key-value (KV) cache. A serving system can reuse these states when a later prompt begins with the same sequence of tokens, called a \emph{shared prefix}. The remaining portion of the prompt is its \emph{suffix}.

Whether reuse improves performance also depends on cache management. GPU memory is limited, and retaining cached states for one call can displace states needed by another. Evicting a prefix before its next use forces recomputation, while retaining context that will not be reused wastes scarce memory. Under KV-cache thrashing, per-request prefill time can grow by $6.8\times$ and generation throughput can drop by up to 66\%~\cite{yuan2026agentic}. Efficient serving therefore requires both exposing reusable prefixes and deciding which cached states deserve memory between calls.

Existing systems exploit prefix sharing and workflow predictability. SGLang's RadixAttention reuses the longest cached prefix matching each incoming prompt and prioritizes requests with longer matches~\cite{zheng2024sglang}. Workflow-aware systems such as KVFlow, Helium, PBKV, and CacheScout~\cite{pan2025kvflow,wadlom2026helium,zheng2026pbkv,zhang2026cachescout} use workflow structure or observed execution patterns to anticipate upcoming calls and guide cache management. KVFlow, for example, uses predicted agent execution order to prioritize retention of fixed prompt prefixes, while assigning varying suffixes the highest eviction priority. This policy treats the suffix as opaque with respect to its components' reuse potential: it does not distinguish persistent shared values from temporary content within that suffix.

\textbf{However, we observe that the dynamic suffix has internal structure: it comprises distinct components with different sharing patterns and lifetimes, a property we term intra-prompt heterogeneity.} A session-wide input, an accumulating history, and a temporary result may appear together in one prompt, yet have very different reuse potential. Some values recur across several calls, some preserve earlier content as they grow, and others cease to be needed after a single iteration. Recognizing this structure creates opportunities to expose sharing hidden by the current prompt order and to distinguish useful cached states from those that can be released.

\begin{figure}[h]
    \centering
    \includegraphics[width=1.0\linewidth]{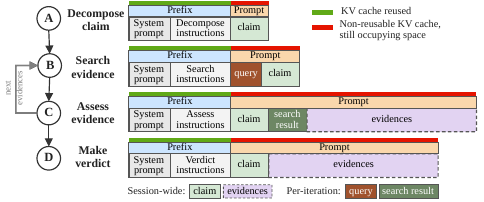}
    \caption{Original prompt layouts for a fact-checking agent. Call-specific instructions precede shared context, limiting prefix reuse across call sites. The claim remains unchanged, evidence accumulates, and queries and individual search results are replaced between iterations.}
    \label{fig:original-view}
    \vspace{-10pt}
\end{figure}

Consider the fact-checking agent in Figure~\ref{fig:original-view}. Given a claim, the agent generates a query (A), searches for evidence (B), and assesses the result against previously collected evidence (C). It repeats the search and assessment until it can produce a verdict (D). In this example, prior systems reuse fixed instruction prefixes consisting of the common system prompt followed by the instructions for decomposition (A), search (B), assessment (C), or verdict generation (D). Their cache retention and eviction policies manage these identified prefixes without distinguishing the sharing patterns and lifetimes of individual components in the dynamic suffix. The claim flows unchanged to all four call sites and remains useful throughout the session. The evidence collection is shared by assessment and verdict calls and preserves accumulated content as it grows. Queries and individual search results are replaced between iterations. These differences remain hidden when the dynamic suffix is treated as opaque.

\begin{figure}[t]
    % \vspace{-8pt}
    \centering
    \includegraphics[width=0.85\linewidth]{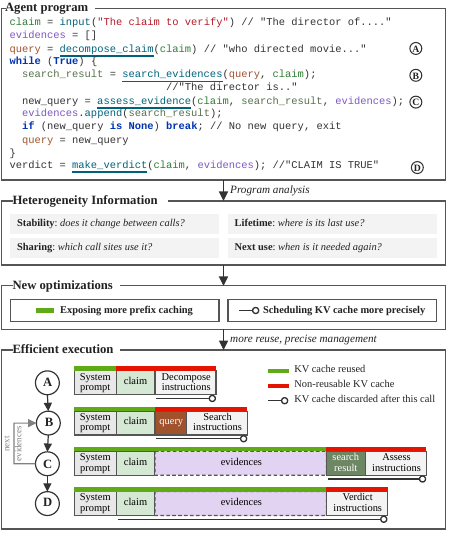}
    \caption{Program analysis recovers sharing, updates, and lifetime information for prompt components. Combined with runtime observations, this information guides coordinated prompt layouts and cache retention. Shared context moves forward, while discard markers identify cached tails that can be released after a call.}
    \label{fig:intra-prompt}
    \vspace{-8pt}
\end{figure}

\textbf{First, this structure enables prompt layouts that expose more prefix reuse.} In Figure~\ref{fig:original-view}, each call site's fixed instructions can be reused across its invocations, but differing instructions before the shared claim prevent prefix reuse across sites. Moving the claim immediately after the common system prompt exposes a shared prefix across all four sites. Placing accumulated evidence before temporary inputs similarly exposes its unchanged token prefix for reuse in later calls. Once these prefixes are exposed, the serving engine's existing prefix caching mechanism can reuse them.
\textbf{Second, this structure enables retention decisions based on future uses.} The claim will be consumed by later calls, whereas cached states for an iteration's temporary tails may never be reused. Figure~\ref{fig:intra-prompt} illustrates retaining shared prefixes while releasing disposable tails. Although a search result's content is appended to the evidence collection, its cached states in the original prompt position need not remain reusable. Component relationships must therefore be interpreted together with prompt order: layout determines which values form reusable prefixes, and lifecycle information helps determine how long those prefixes are worth retaining.

\textbf{Our key insight is that the sharing patterns and lifetimes of program values provide actionable information for exposing additional prefix reuse and retaining cached states according to their future uses. More importantly, we observe that this sharing and lifetime information is already encoded in the agent program itself, making it recoverable through compiler analysis.} Frameworks such as byLLM, NOOA, and DSPy~\cite{dantanarayana2025mtp,furgale2026nooa,khattab2024dspy} generate prompts from typed LLM function declarations and argument values, establishing a correspondence between prompt components and program variables. In Figure~\ref{fig:intra-prompt}, dataflow connects the unchanged claim to all four call sites, the append operation exposes growing evidence, and assignments inside the loop identify replaced queries. Control flow analysis identifies possible future consumers. Recovering these relationships makes it possible to coordinate layouts across calls and assess the future usefulness of cached prefixes, even before those future prompts arrive.

Turning this insight into efficient serving requires addressing three challenges. \textbf{First, program structure alone does not determine actual reuse.} Static analysis reveals possible sharing, updates, and future consumers, but execution paths, component sizes, and inter-call delays vary at runtime. \textbf{Second, exposing prefix reuse requires coordination across call sites.} Independently ordering each call's components can place the same value behind different preceding contexts, preventing its cached states from being shared. \textbf{Third, potential reuse must be weighed against current memory demand.} Retaining or prefetching a prefix for a future call consumes memory that other requests could use, while premature eviction forces recomputation.

We present \sysname, a co-design of the compiler and serving runtime that addresses these challenges through a common description of prompt components and their uses. To connect program structure with actual execution, the compiler produces component metadata and a workflow graph, which the runtime supplements with observations to estimate future reuse and its timing. Using this information, \emph{adaptive prompt re-layout} coordinates component ordering across related calls to expose longer shared prefixes. \emph{Proactive KV-cache scheduling} then uses component lifetimes and predicted reuse to prioritize retention and eviction. It bounds retention with a fixed memory budget and prefetches only into available space, reclaiming disposable cached tails when needed. Together, these mechanisms connect the analysis of program values to both the construction of reusable prefixes and the allocation of memory to preserve them.

We make the following contributions:
\begin{enumerate*}[label=\textbf{(\roman*)}]
    \item We identify intra-prompt heterogeneity and the opportunities it creates to expose additional prefix reuse and distinguish the retention needs of prompt components.
    \item We develop compiler analysis to recover component sharing and lifetimes from agent programs, and combine this information with runtime observations to coordinate prompt layout and KV-cache management.
    \item We implement \sysname and evaluate it on four agent workloads. It achieves geometric-mean speedups of 1.45$\times$ in average job completion time (JCT) and 1.49$\times$ in throughput over SGLang, with maxima of 1.99$\times$ and 1.97$\times$, respectively.
\end{enumerate*}

\section{Background and Motivation}
\label{sec:bg_mt}
\para{Prefix caching and KV-cache tiering}
Serving engines such as SGLang~\cite{zheng2024sglang} and vLLM~\cite{kwon2023pagedattention} reuse the longest matching prefix, computing only the remaining tokens. Tiered caching systems such as CachedAttention~\cite{gao2024cachedattention}, Mooncake~\cite{qin2025mooncake}, and SGLang HiCache~\cite{sglang2026hicache} extend the KV-cache beyond device memory with host memory and optionally disk storage to improve capacity. However, off-device KV states must be transferred back to GPU memory before use.

\para{KV-cache management for LLM agents}
Agent serving systems use recurring patterns in agentic workloads to manage the KV-cache. KVFlow~\cite{pan2025kvflow} derives each agent's steps-to-execution from the workflow to guide eviction and prefetching of each agent's prompt prefix; Helium~\cite{wadlom2026helium} analyzes the workflow to order LLM calls and prewarm static prefixes; Learned Prefix Caching~\cite{yang2025lpc}, PBKV~\cite{zheng2026pbkv} and CacheScout~\cite{zhang2026cachescout} predict upcoming agent invocations with trained models to decide which caches to retain. KVTether~\cite{fu2026kvtether} captures runtime lifecycle semantics from agent harness signals to preferentially preserve live-but-idle KV-cache. They operate the KV-cache on the identified prefix. Some systems, such as Pythia~\cite{yu2026pythia}, prefill the predicted next invocations to proactively prepare KV-cache that does not exist yet. 

\para{Frameworks with LLM calls as typed functions}
A class of agent frameworks exposes LLM invocations through typed programming interfaces, such as byLLM's meaning-typed programming (MTP) abstraction~\cite{dantanarayana2025mtp}, DSPy's Signatures and Modules~\cite{khattab2024dspy}, and NOOA's typed Python methods~\cite{furgale2026nooa}. The developer declares a function or signature describing the desired behavior, inputs, and return type, and invokes it like an ordinary call within the program's control flow. At each invocation, the framework runtime lowers the declaration and its argument values into final prompts.

\begin{figure}[t]
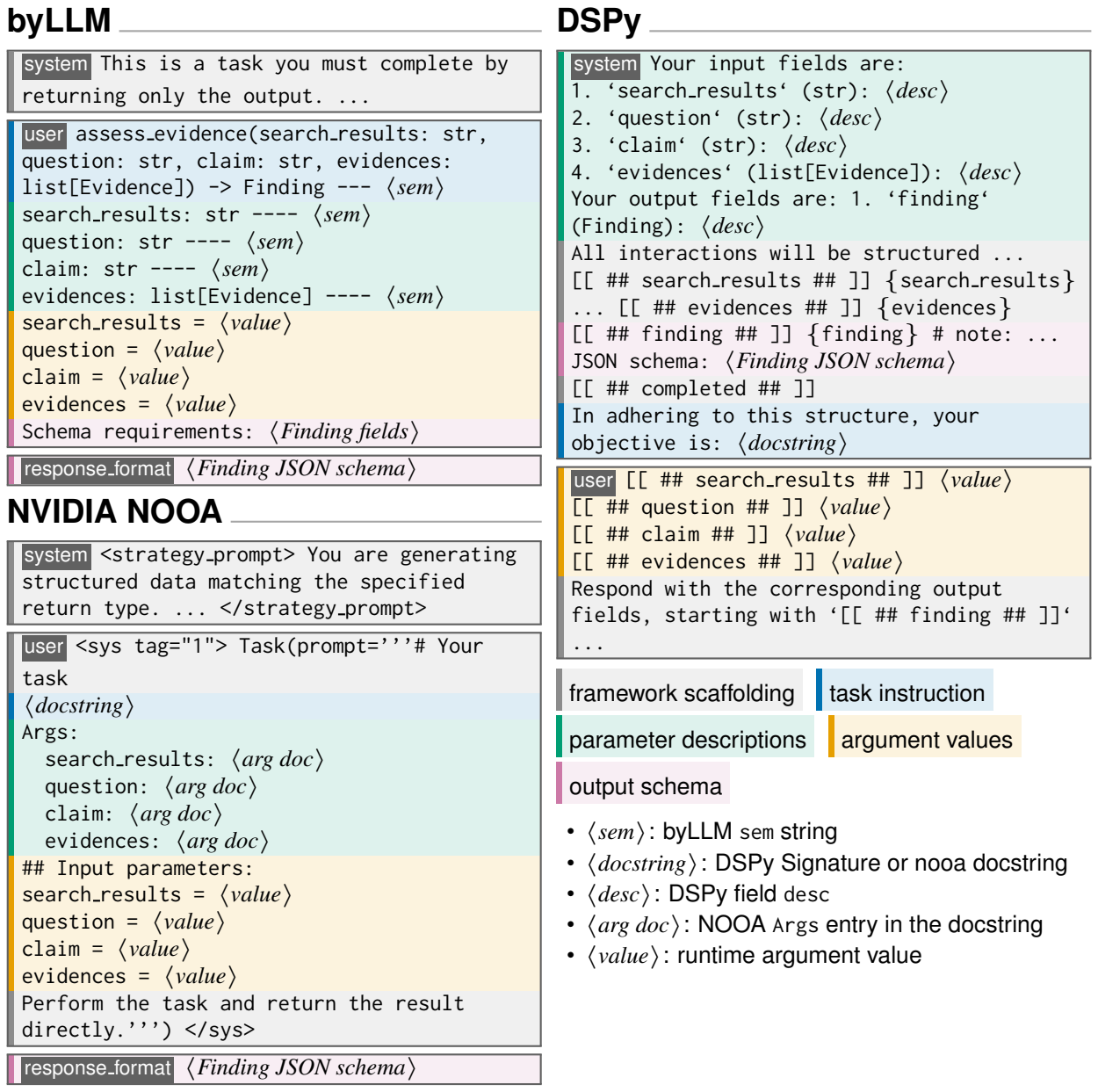

    % \vspace{-8pt}
  \centering
  \definecolor{csTask}{HTML}{0072B2}
  \definecolor{csDesc}{HTML}{009E73}
  \definecolor{csVal}{HTML}{E69F00}
  \definecolor{csSchema}{HTML}{CC79A7}
  \definecolor{csScaf}{gray}{0.55}
  \newlength{\csbar}\setlength{\csbar}{1.2pt}
  \newcommand{\csfont}{\fontsize{5}{5.8}\selectfont}
  \newcommand{\ph}[1]{{\normalfont\itshape$\langle$#1$\rangle$}}
  % \csseg{color}{text}: one component, a tinted band with a bar in its color.
  \newcommand{\csrolechip}{\ifcsfirst{\fboxsep=0.6pt\colorbox{black!62}{\sffamily\fontsize{4.3}{4.3}\selectfont\color{white}\strut\csrole}}\hspace{2pt}\global\csfirstfalse\fi}
  \newcommand{\csseg}[2]{\par\noindent
    \sbox0{\parbox{\dimexpr\linewidth-\csbar-3pt\relax}{\raggedright\hyphenpenalty=10000\strut\csrolechip#2\strut}}%
    {\fboxsep=0pt\colorbox{#1!13}{{\color{#1}\vrule width\csbar height\ht0 depth\dp0}\hspace{1.6pt}\usebox0\hspace{1.4pt}}}\par}
  \newcommand{\csscaf}[1]{\csseg{csScaf}{#1}}
  % \csmsg{role}{segments}: one message as a card, its role on a tab.
  % The role sits as a chip at the start of the card's first line instead of on a tab above it.
  \newif\ifcsfirst
  \newcommand{\csrole}{}
  \newcommand{\csmsg}[2]{\par\nointerlineskip\noindent
    \renewcommand{\csrole}{#1}\csfirsttrue
    \vbox{\offinterlineskip
      \hbox{\fboxsep=0pt\fboxrule=0.4pt\fcolorbox{black!62}{white}{\parbox{\dimexpr\linewidth-2\fboxrule\relax}{\ttfamily\csfont\offinterlineskip#2}}}}%
    \par\nointerlineskip\vspace{1.4pt}}
  \newcommand{\cshead}[1]{\par\nointerlineskip\noindent{\sffamily\scriptsize\bfseries\strut #1}\,{\color{black!35}\hrulefill}\par\nointerlineskip\vspace{1.4pt}}
  \newcommand{\cskey}[2]{{\fboxsep=0pt\colorbox{#1!13}{{\color{#1}\vrule width\csbar height 1.55ex depth 0.45ex}\hspace{1.6pt}{\sffamily\fontsize{5}{6}\selectfont#2}\hspace{1.6pt}}}}
  \begin{minipage}[t]{0.488\linewidth}
    \cshead{byLLM}
    \csmsg{system}{%
      \csscaf{This is a task you must complete by returning only the output. ...}}
    \csmsg{user}{%
      \csseg{csTask}{assess\_evidence(search\_results: str, question: str, claim: str, evidences: list[Evidence]) -> Finding --- \ph{sem}}
      \csseg{csDesc}{search\_results: str ---- \ph{sem}\newline question: str ---- \ph{sem}\newline claim: str ---- \ph{sem}\newline evidences: list[Evidence] ---- \ph{sem}}
      \csseg{csVal}{search\_results = \ph{value}\newline question = \ph{value}\newline claim = \ph{value}\newline evidences = \ph{value}}
      \csseg{csSchema}{Schema requirements: \ph{Finding fields}}}
    \csmsg{response\_format}{%
      \csseg{csSchema}{\ph{Finding JSON schema}}}
    \vspace{0.8pt}
    \cshead{NVIDIA NOOA}
    \csmsg{system}{%
      \csscaf{<strategy\_prompt> You are generating structured data matching the specified return type. ... </strategy\_prompt>}}
    \csmsg{user}{%
      \csscaf{<sys tag="1"> Task(prompt='''\# Your task}
      \csseg{csTask}{\ph{docstring}}
      \csseg{csDesc}{Args:\newline \hspace*{1em}search\_results: \ph{arg doc}\newline \hspace*{1em}question: \ph{arg doc}\newline \hspace*{1em}claim: \ph{arg doc}\newline \hspace*{1em}evidences: \ph{arg doc}}
      \csseg{csVal}{\#\# Input parameters:\newline search\_results = \ph{value}\newline question = \ph{value}\newline claim = \ph{value}\newline evidences = \ph{value}}
      \csscaf{Perform the task and return the result directly.''') </sys>}}
    \csmsg{response\_format}{%
      \csseg{csSchema}{\ph{Finding JSON schema}}}
  \end{minipage}\hfill
  \begin{minipage}[t]{0.488\linewidth}
    \cshead{DSPy}
    \csmsg{system}{%
      \csseg{csDesc}{Your input fields are:\newline 1. `search\_results` (str): \ph{desc}\newline 2. `question` (str): \ph{desc}\newline 3. `claim` (str): \ph{desc}\newline 4. `evidences` (list[Evidence]): \ph{desc}\newline Your output fields are: 1. `finding` (Finding): \ph{desc}}
      \csscaf{All interactions will be structured ...\newline [[ \#\# search\_results \#\# ]] \{search\_results\}\newline ... [[ \#\# evidences \#\# ]] \{evidences\}}
      \csseg{csSchema}{[[ \#\# finding \#\# ]] \{finding\} \# note: ... JSON schema: \ph{Finding JSON schema}}
      \csscaf{[[ \#\# completed \#\# ]]}
      \csseg{csTask}{In adhering to this structure, your objective is: \ph{docstring}}}
    \csmsg{user}{%
      \csseg{csVal}{[[ \#\# search\_results \#\# ]] \ph{value}\newline [[ \#\# question \#\# ]] \ph{value}\newline [[ \#\# claim \#\# ]] \ph{value}\newline [[ \#\# evidences \#\# ]] \ph{value}}
      \csscaf{Respond with the corresponding output fields, starting with `[[ \#\# finding \#\# ]]` ...}}
    \vspace{0.6pt}
    {\raggedright\noindent\baselineskip=7.5pt
    \cskey{csScaf}{framework scaffolding}\hspace{3pt}\cskey{csTask}{task instruction}\hspace{3pt}\cskey{csDesc}{parameter descriptions}\hspace{3pt}\cskey{csVal}{argument values}\hspace{3pt}\cskey{csSchema}{output schema}\par}
    \vspace{3pt}
    {\raggedright\sffamily\fontsize{5}{6.6}\selectfont
    \begin{itemize}[nosep, leftmargin=6pt, labelsep=2pt, label=\textbullet, topsep=0pt, itemsep=0pt]
      \item \ph{sem}: byLLM \texttt{sem} string
      \item \ph{docstring}: DSPy Signature or nooa docstring
      \item \ph{desc}: DSPy field \texttt{desc}
      \item \ph{arg doc}: NOOA \texttt{Args} entry in the docstring
      \item \ph{value}: runtime argument value
    \end{itemize}\par}
  \end{minipage}
  % mlsys2025.sty hard-codes 10pt above captions; trimming it lets §2.5 start on this page.
  \vspace{-6pt}
  \caption{Prompts lowered by three frameworks for the same call share five kinds of components; only their order differs.}
  \label{fig:common-structure}
    \vspace{-8pt}
\end{figure}

Although these agent frameworks differ in surface syntax, they lower LLM invocations in the same way: the intermediate representation is lowered with the runtime argument values to produce the prompt. As shown in Figure~\ref{fig:common-structure}, these frameworks emit the same five kinds of components: framework scaffolding, task instruction, parameter descriptions, argument values, and output schema. The difference is their relative positions. 

\emph{Prompt-layout Opportunity:}
By default, component order follows framework templates and argument declarations. These layouts can be adjusted to expose more prefix reuse. For example, DSPy's static adapter~\cite{dspy2026chatadapter} lets users tweak the lowering process to put static prompt components at the front. NOOA~\cite{furgale2026nooa} lets developers label arguments as static prefix, append-only history or dynamic suffix so that the framework can order them for better cache reuse.

\begin{figure}[t]
    % \vspace{-8pt}
    \centering
    \includegraphics[width=0.85\columnwidth]{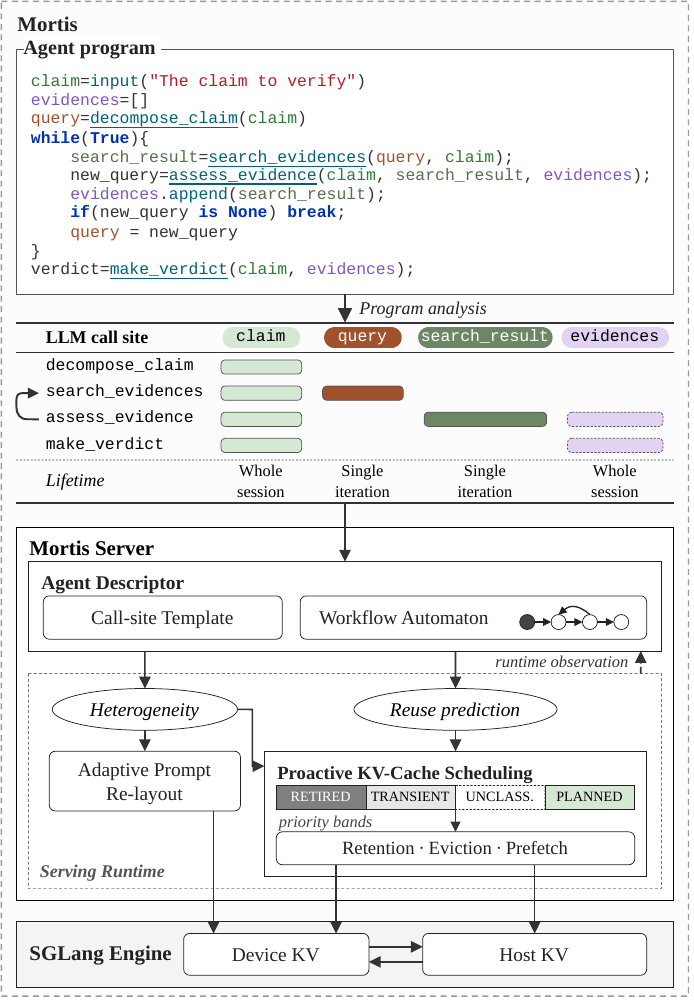}
    \caption{Architecture overview of \sysname. Program analysis and runtime observations guide prompt re-layout and KV-cache management.}
    \label{fig:design_overview}
    \vspace{-8pt}
\end{figure}

\para{Intra-prompt heterogeneity in programs}
Prompts contain components with different sharing patterns and lifetimes. In frameworks with typed LLM calls, components generated from arguments correspond to program variables. Program analysis can recover which values are shared, extended, or replaced, and identify possible future uses.

\emph{Cache-management Opportunity:}
Prior serving policies fail to fully exploit the reuse in agent workloads because they ignore the component-level structure of prompts. Manual annotations mitigating this gap like NOOA's labels restate properties that already exist in the program, and may become stale once the code is updated. Instead, reasoning about the program itself can precisely guide better prefix reuse and KV-cache retention decisions by discovering sharing and lifetime information.

\para{From Deficiencies to Design}
Ignoring intra-prompt heterogeneity costs a serving system in two ways. First, shared tokens lose prefix caching: once a differing token appears ahead of them, their KV-cache can no longer be matched. Second, KV-cache retention loses precision: the system sees only a matched prefix and cannot manage entries by their individual reuse. Both costs are substantial. Among the agents we evaluated in \S\ref{sec:eval}, 51--92\% of shared tokens miss the prefix cache because a differing token precedes them, and about 20\% of served prompt tokens are used only once within a session.  The layout freedom and the analyzability of intra-prompt heterogeneity address these deficiencies directly, upon which \sysname is built.

\section{Design}

\label{sec:design}
In this section, we present the two-tier compiler-serving co-design of \sysname. As shown in Figure~\ref{fig:design_overview}, the \emph{agent descriptor} is generated by combining analysis of the agent program with observation of runtime behaviors. The server side then exploits the derived descriptor to schedule the KV-cache efficiently, using \emph{adaptive prompt re-layout} (\S\ref{sec:relayout}) and \emph{proactive KV-cache scheduling} (\S\ref{sec:Proactive KV-Cache Scheduling}).

\subsection{Agent Descriptor}
The \emph{agent descriptor} contains the metadata for scheduling. Static analysis provides its skeleton, and the server refines the weights with runtime observations. An agent descriptor contains two parts: 
\begin{enumerate*}[label=(\roman*)]
    \item \textbf{Call-site Template:} It provides a heuristic for each component's lifetime and sharing pattern.
    \item \textbf{Workflow Automaton:} It contains the program-level structure of an agent, including workflow dependencies, execution time estimation and transition probabilities.
\end{enumerate*} 

\subsubsection{Call-site Template}
\label{sec:template}

Prompts can be decomposed into heterogeneous components. We classify these components into the following two categories:
\begin{enumerate*}[label=(\roman*)]
    \item \textbf{Invariant Specifications:} the part of prompt that is fixed once the program is compiled. They are rendered from the compiler's own intermediate representation and stay constant.
    \item \textbf{Dynamic Bindings:} the argument values of LLM functions, which are determined at runtime.
\end{enumerate*}

We define the call-site template as $\mathcal{T}_c=\langle \mathcal{I}_c,\ \mathcal{B}_c,\ \mathcal{H}_c \rangle$, where $\mathcal{I}_c$ denotes the invariant specifications, $\mathcal{B}_c$ the dynamic bindings, and $\mathcal{H}_c$ the heterogeneity annotation over those bindings.

\paragraph{Heterogeneity annotation}
For each dynamic binding, $\mathcal{H}_c$ records how its value is shared and how long that value stays valid, which we represent as $\mathcal{E}$ and $\mathcal{S}$ respectively. $\mathcal{S}$ is the set of program regions that overwrite a binding, and $\mathcal{E}$ classifies a binding into five classes (Table~\ref{tab:evolution_classes}) by reusability.
\begin{table}[h]
    \centering
    \small
    \caption{Evolution classes of a dynamic binding, from the most to the least reusable.}
    \label{tab:evolution_classes}
    \setlength{\tabcolsep}{4pt}
    \begin{tabular}{@{}lp{0.79\columnwidth}@{}}
        \toprule
        Class & Value relationship \\
        \midrule
        CONST    & Fixed by the code and identical at every call. \\
        COPY     & Same as the value another call-site carries. \\
        RESP     & Comes from the reply produced by another call-site. \\
        EXTEND   & Append-only context; the previous value remains a prefix of the new one. \\
        VOLATILE & Freshly produced at each call, or its origin lies outside the analysis boundary. \\
        \bottomrule
    \end{tabular}
\end{table}
% We write $\rho$ for the rank of a class in the order above, which expresses how much of a binding's bytes an earlier call can already supply. 
\subsubsection{Workflow Automaton}
\label{sec:TPFM}
We abstract an agent's workflow as a time-annotated probabilistic finite-state automaton. We formulate it as: 
\[
\mathcal{A} = (Q, P, q_0, F, D^{\mathrm{gap}}),
\]
where:
\begin{enumerate*}[label=(\roman*)]
    \item $Q$ is the set of states. Each state $q\in Q$ is labeled by an LLM call-site;
    \item $P$ denotes the transition probabilities between states;
    \item $q_0 \in Q$ is the entrance call-site;
    \item $F$ specifies the termination probability of the agent at each state;
    \item $D^{\mathrm{gap}}(q,q')$ is the inter-call gap distribution between the completion of call-site $q$ and the invocation of its successor $q'$.
\end{enumerate*}

\subsection{Constructing the Agent Descriptor}
\label{sec:analysis}

\para{Call-site graph construction}
The compiler identifies LLM call-sites from the typed AST. It then builds the interprocedural control-flow graph (CFG) of the agent and projects it onto these call-sites. An edge $(q, q')$ is added whenever $q'$ is directly reachable in one hop from $q$, while non-LLM function call-sites are contracted. If a call target cannot be resolved statically, the compiler conservatively adds edges to all candidate targets. This projected graph forms the unweighted structure of $\mathcal{A}$. Finally, the compiler applies Bourdoncle's weak topological ordering~\cite{bourdoncle1993efficient} to recover the loop nesting structure from the cycles in $\mathcal{A}$.

\para{Separating $\mathcal{I}_c$ and $\mathcal{B}_c$}
$\mathcal{I}_c$ and $\mathcal{B}_c$ are obtained by running the lowering process that generates actual prompts. The compiler runs the framework's lowering pass ahead of time using symbolic placeholders for the arguments to produce the static portion of the runtime prompt ($\mathcal{I}_c$) while leaving one labeled placeholder per parameter to represent $\mathcal{B}_c$.

\para{Recovering evolution classes ($\mathcal{E}$)}
For each binding $b \in \mathcal{B}_c$, the compiler computes the reaching definitions of its underlying variable across the CFG (including back edges, with local definitions shadowing earlier ones). Each reaching definition is classified by its right-hand side:
\begin{enumerate*}[label=(\roman*)]
    \item literals and module-level constants are classified as \textbf{CONST};
    \item variables passed unmodified from an earlier call-site $c'$ are classified as \textbf{COPY}($c'$);
    \item values assigned from the return value of another call-site $c'$ (traced via def--use chains) are classified as \textbf{RESP}($c'$);
    \item prefix-preserving updates, like string/list concatenation ($v \leftarrow v + \Delta$), are classified as \textbf{EXTEND};
    \item all other definitions are defaulted to \textbf{VOLATILE}, such as external input.
\end{enumerate*}

If different CFG paths yield conflicting evolution classes for $b$, the compiler falls back to \textbf{VOLATILE}. To let the runtime resolve this conservativeness once the actual transition $(q, q')$ is known, the compiler attaches transition-specific metadata to each edge $(q, q')$. Specifically, it records which and how the variables are modified along $(q,q')$, and attaches the exact evolution class for each binding at $q'$. For example, a binding can be reduced to \textbf{CONST} rather than \textbf{VOLATILE} if the exact edge assigns it a literal.

\para{Recovering lifetime scopes ($\mathcal{S}$)}
The compiler records $\mathcal{S}(b)$ as the set of procedures (or loop bodies) that overwrite the binding $b$'s underlying variable. When the variable is assigned once and remains valid for the entire session, $\mathcal{S}(b)=\emptyset$. Whenever execution enters a procedure in $\mathcal{S}(b)$, the variable is overwritten and its previously cached KV bytes become dead.

\subsection{Adaptive Prompt Re-layout}
\label{sec:relayout}
\begin{algorithm}[t]
    \caption{Layout decision }
    \label{alg:layout}
    \small\setlength{\lineskiplimit}{-3pt}
    \begin{algorithmic}[1]
    \algtext*{EndFor}\algtext*{EndIf}\algtext*{EndWhile}
    \Require Descriptor $(\mathcal{A},\{\mathcal{T}_q\}_{q\in Q})$, switch \textit{pin}, lengths $\ell$
    \Ensure Layouts $\{\mathcal{L}_q\}_{q\in Q}$
    \State $B_{\mathrm{sh}} \gets \{b : |\{q\in Q : b\in\mathcal{B}_q\}|>1\}$\label{alg:layout:sh-begin}\Comment{shared bindings}
    \State $G \gets(B_{\mathrm{sh}},\emptyset)$ \Comment{precedence graph}
    \ForAll{$(q,q')\in\Call{Edges}{\mathcal{A}}$}
        \State $P_1,P_2,P_3 \gets \emptyset$
        \Statex \Comment{Groups: unchanged, append, rewritten/one-sided}
        \ForAll{$b \in B_{\mathrm{sh}}\cap(\mathcal{B}_q\cup\mathcal{B}_{q'})$}
            \State $P_{\tau(b)} \gets P_{\tau(b)}\cup\{b\}$ \Comment{$\tau(b)$: group assignment of $b$}
        \EndFor
        \State $E(G) \gets E(G)\cup\{(b,b')\in P_i\times P_k : i<k\}$
    \EndFor
    \State $\mathcal{C} \gets \Call{Tarjan}{G}$ \Comment{SCCs of $G$, sets of bindings}
    \State $\hat{E} \gets \{(C,C')\in\mathcal{C}^2 : C\neq C',\ (C\times C')\cap E(G)\neq\emptyset\}$
    \State $\hat{G} \gets (\mathcal{C},\hat{E})$ \Comment{condensation graph of $G$}
    \State $j(b) \gets (-\ell(b),\textit{depth}(b)),\ \forall b\in B_{\mathrm{sh}}$
    \State $J(C) \gets \min_{b\in C} j(b),\ \forall C\in V(\hat{G})$
    \State $L \gets [\,]$
    \While{$V(\hat{G})\neq\emptyset$} \Comment{topological sort}
        \State $R \gets \{C\in V(\hat{G}) : \textit{indeg}(C)=0\}$
        \State $C^{*} \gets \arg\min_{C\in R} J(C)$
        \State $L \gets L \concat \Call{Sort}{C^{*},j}$;\quad $\hat{G} \gets \hat{G}-C^{*}$ \label{alg:layout:sh-end}
    \EndWhile
    \ForAll{$q\in Q$}
        \State $\sigma_q \gets [\,b\in L : b\in\mathcal{B}_q\,] \concat \Call{Sort}{\mathcal{B}_q\setminus B_{\mathrm{sh}},K}$ \label{alg:layout:pv}
        \State $s_q \gets \Call{Spec}{\mathcal{I}_q}$ \Comment{instruction specification of $q$} \label{alg:layout:spec-begin}
        \If{$\neg\textit{pin}\land(\mathcal{B}_q\cap B_{\mathrm{sh}}\neq\emptyset)$}
            \State $\mathcal{L}_q \gets \sigma_q \concat s_q$ \Comment{spec last}
        \Else
            \State $\mathcal{L}_q \gets s_q \concat \sigma_q$ \Comment{spec first}
        \EndIf
        \label{alg:layout:spec-end}
    \EndFor
    \State \Return $\{\mathcal{L}_q\}_{q\in Q}$
    \end{algorithmic}
\end{algorithm}

\subsubsection{Movable Components}
First, the dynamic bindings $\mathcal{B}_c$ can be permuted freely. The compiler lowers each argument into a self-contained component labeled by its parameter name, and the framework places these components in declaration order only because that is the order of the argument list; re-layout may choose any order of the bindings.

Second, the instruction specification in $\mathcal{I}_c$ is movable. $\mathcal{I}_c$ consists of the system prompt, the response schema, the tool schema, and the instruction specification (such as task descriptions and parameter descriptions). The system prompt stays universally constant and should be placed at the very beginning, and the schemas are handled by the constrained decoding engine. The instruction specification can be put just after the system prompt or at the end of the prompt.

\subsubsection{Layout Decision}
\label{sec:layout_decision}

\sysname decides the layout $\mathcal{L}_q$ for each call-site $q$ by ranking the reusability of bindings (Algorithm~\ref{alg:layout}). It places shared bindings ahead of private bindings and prioritizes stable bindings within both categories to maximize prefix cache efficiency. 

\paragraph{Sorting shared bindings {\normalfont(lines~\ref{alg:layout:sh-begin}--\ref{alg:layout:sh-end})}} Shared bindings are those that are carried by multiple call-sites. For every transition $(q,q')$, we identify how each shared binding evolves. We categorize their behaviors into three sharing pattern groups, ordered as follows:
\begin{enumerate*}[label=(\textbf{\roman*})]
    \item \textbf{unchanged:} Both call-sites carry it and its value does not change between the two calls.
    \item \textbf{append:} Both call-sites carry it and the program only appends to its value.
    \item \textbf{rewritten/one-sided:} The program rewrites it between the two calls, or only one call-site carries it.
\end{enumerate*}

\sysname builds the precedence graph of all shared bindings based on the groups they belong to on each transition edge, condenses it into a DAG using Tarjan's algorithm, and finally performs a topological sort on the DAG. We key each SCC by the smallest $\textbf{(-\textit{length}, \textit{depth})}$ among its bindings, where:
\begin{enumerate*}[label=\textbf{(\roman*)}]
    \item \textit{depth}: the number of transitions from the entrance to the first call-site that carries $b$;
    \item \textit{length}: the token length of the value of $b$, observed by the server.
\end{enumerate*}
The SCC with the smallest key goes first when multiple SCCs have zero in-degree. Each SCC is then expanded into the list of its bindings sorted also by $\textbf{(-\textit{length}, \textit{depth})}$.

\paragraph{Sorting private bindings {\normalfont(line~\ref{alg:layout:pv})}} The bindings carried by a single call-site are sorted in ascending lexicographic order of:
\[
K(b)=\big(\textit{scoped},\ \rho,\ \textit{depth},\ -\textit{length}\big),
\]
where:
\begin{enumerate*}[label=\textbf{(\roman*)}]
    \item \textit{scoped}: the lifetime of $b$; 0 if it is session-wide ($\mathcal{S}=\emptyset$) and 1 otherwise;
    \item $\rho$: the stability rank for $\mathcal{E}$ in Table~\ref{tab:evolution_classes}, from 1 for CONST to 5 for VOLATILE.
\end{enumerate*}

\paragraph{Placing the specification {\normalfont(lines~\ref{alg:layout:spec-begin}--\ref{alg:layout:spec-end})}} 
Placing the \emph{instruction specification} at the end gives a longer common prefix. \sysname places the specification at the end for each call-site that carries a shared binding. A switch keeps the specification at the front when the developer requires it.

\subsection{Proactive KV-cache Scheduling}
\label{sec:Proactive KV-Cache Scheduling}
% \sysname's KV-cache scheduling proactively arranges the placement of KV-cache entries. Using the agent descriptor, the server decides operations for every served KV-cache sequence.

\subsubsection{Reusable Head and Volatile Tail}
\label{sec:identify_reusable_kv}
A served prompt naturally divides into two segments: a reusable head and a volatile tail. The reusable head consists of two parts: a \emph{static prefix} and a \emph{session-shared extension}:
\begin{enumerate*}[label=(\roman*)]
    \item \textbf{Static Prefix} is the prefix that does not change, including $\mathcal{I}_c$ and the CONST bindings that re-layout has placed behind $\mathcal{I}_c$.
    \item \textbf{Session-shared Extension} consists of the bindings after the static prefix whose values recur unchanged in a later call of the same session.
\end{enumerate*}

\subsubsection{Next-Call Prediction}
\sysname predicts three aspects of the next call:
\begin{enumerate*}[label=(\roman*)]
    \item The probability of the arrival.
    \item When the call will arrive.
    \item Which prefix of its prompt can be speculatively constructed.
\end{enumerate*}

\paragraph{Arrival probability}
Each call-site $q$ can be located by a chain of loops enclosing it; for each call-site's invocation, we mark it with a tuple $\zeta=\left(i_1, i_2, \dots, i_m\right)$, where $i_k, k \in \left[1, m\right]$ is the current iteration index of the $k$-th enclosing loop, from outermost to innermost. 

\sysname estimates the probability of the next call by a loop-tagged table. We define the table as $\mathcal{P}(q' \mid q, \zeta)$ and $\mathcal{F}(q, \zeta)$, where $\mathcal{P}$ is the probability that call-site $q$ in context $\zeta$ is followed by call-site $q'$; $\mathcal{F}$ stands for the probability that the session ends at call-site $q$ under context $\zeta$. With an empty $\zeta$ (no loop context), $\mathcal{P}$ and $\mathcal{F}$ reduce to the base transition and termination probabilities $P$ and $F$ of the automaton $\mathcal{A}$.

The table is approximated and updated by frequency counts observed at runtime. Let $q$ and $\zeta$ denote the call-site and loop context of the preceding call in the same session. When a call at $q'$ arrives, the server increments the count of outcome $q'$ in row $(q,\zeta)$; when the session closes at $q$, it increments the termination count in row $(q,\zeta)$. Each observation is also counted in every row $(q,\zeta_k)$ with $\zeta_k=(i_{m-k+1},\dots,i_m)$, $k=0,\dots,m$.

After a call at $q$ lands, we compute the probability of each outcome $o$ by Witten--Bell interpolation~\cite{witten1991zero} from the empty context $\zeta_0$ to the full context $\zeta_m=\zeta$. The outcome $o$ ranges over $\mathrm{succ}(q)$, the static successors of $q$ plus a virtual outcome $\mathrm{end}$ for session termination:
\begin{align*}
\hat{P}_k(o) &= \lambda_k\frac{c_k(o)}{n_k}+(1-\lambda_k)\hat{P}_{k-1}(o),\quad k=0,\dots,m, \\
\hat{P}_{-1}(o) &= \frac{1}{|\mathrm{succ}(q)|},\qquad
\lambda_k=\begin{cases}\dfrac{n_k}{n_k+d_k} & n_k>0,\\[4pt] 0 & n_k=0,\end{cases}
\end{align*}
$c_k(o)$ is how many times outcome $o$ followed a call at $q$ in context $\zeta_k$, $n_k=\sum_o c_k(o)$ is the total number of observations in that row, and $d_k$ is the number of distinct outcomes with $c_k(o)>0$. Then $\mathcal{P}(q'\mid q,\zeta)=\hat{P}_m(q')$ and $\mathcal{F}(q,\zeta)=\hat{P}_m(\mathrm{end})$. A context with no observation is skipped. The weight $\lambda_k$ grows with $n_k$, so a frequently observed context dominates the prediction, while a rarely observed one defers to its shorter contexts.

\paragraph{Arrival time}
The time of arrival is conservatively approximated based on recorded statistics. The workflow automaton records $D^{\mathrm{gap}}(q,q')$ per transition: each time a call at $q'$ arrives, the server appends the interval between the reply of the preceding call at $q$ and this arrival. The interval is the agent's own time spent in its code between two calls. We estimate the time gap to the next call as:
\[
\hat{g}= \max \{t_{50}-\tfrac{1}{2}(t_{90}-t_{50}), 0 \}
\]
where $t_{50}$ and $t_{90}$ are the 50th and 90th percentiles of the gap samples on the transition $(q,q')$; The second term ensures that a transition with a more dispersed gap distribution receives a tighter estimate. 

% the predicted call is then expected to arrive $\hat{g}$ after the reply of the current call. 

\paragraph{Prefix reconstruction}
\sysname constructs the prefix of call-site $q'$ for matching KV-cache entries to schedule. Since $\mathcal{H}_c$ records the relationships among arguments, \sysname builds the prefix in layout $\mathcal{L}_{q'}$ using the argument values observed in the current session. The construction stops at the first argument with an unknown value.

\subsubsection{KV-cache Eviction Using Lifecycle Bands}
\paragraph{Bands}
We categorize KV-cache entries into four bands, from first evicted to last:
\begin{enumerate*}[label=(\arabic*)]
    \item \textbf{RETIRED:} the volatile tails of prompts belonging to completed sessions. 
    \item \textbf{TRANSIENT:} the volatile tail of a served prompt of a live session. 
    \item \textbf{UNCLASSIFIED:} entries that analysis cannot categorize. They keep the engine's default treatment.
    \item \textbf{PLANNED:} prefixes that a predicted call will reuse. 
\end{enumerate*}

\paragraph{Band transitions}
Bands are assigned at three moments of a session's lifecycle.
\begin{enumerate*}[label=(\arabic*)]
    \item \textbf{When a served prompt enters the cache}, \sysname splits it at its \emph{reusable head} (\S\ref{sec:identify_reusable_kv}). The volatile tail and the reply generated after it become \emph{TRANSIENT}; the head stays \emph{UNCLASSIFIED} until the next prediction refresh finds a successor call that starts with it, which tries to promote it to \emph{PLANNED}.
    \item \textbf{When a later call arrives}, the transition it takes may rewrite bindings that earlier heads carried. Each affected head is cut back to the prefix that survives the rewrite, and the bytes past it are demoted to \emph{TRANSIENT}.
    \item \textbf{When the session ends}, the volatile tails of every prompt it served drop to \emph{RETIRED}. Their heads stay \emph{UNCLASSIFIED} and age out by recency, since other sessions of the same program may share them.
\end{enumerate*}

\paragraph{Band-ordered eviction}
Eviction empties the lowest band first and falls back to recency within the three lower bands. The radix tree's eviction heap is keyed by band, then by recency. The host tier keeps LRU eviction, except that \emph{RETIRED} entries go first. Additionally, KV-cache entries marked as \emph{TRANSIENT} are not admitted into the host tier as they will not be reused.

\subsubsection{KV-cache Retention and Prefetch}
% \sysname protects the required KV-cache entries based on prediction. Each prediction is summarized into a tuple of reconstructed prefix, arrival probability and estimated deadline:
A future reuse prediction is summarized as the following triple:
\[
\textit{Prediction}=\left(\pi_{q'},\ \mathcal{P}(q'\mid q,\zeta),\ t_{\mathrm{reply}}+\hat{g}\right)
\]
where $\pi_{q'}$ is the prefix reconstructed for $q'$ and $t_{\mathrm{reply}}$ is the time the current call replied. Predictions are refreshed at every call, and are dropped when the predicted calls arrive or their predicted arrival time passes. 

\sysname acts on this triple in two ways: \emph{Retention} keeps $\pi_{q'}$ from leaving the device, and \emph{prefetch} restores the part of $\pi_{q'}$ that has already been demoted to the host.

\paragraph{Retention}
\sysname ranks the predicted reusable KV-cache entries by the benefit of keeping $\pi_{q'}$ resident at every refresh:
\[
\mathcal{D}(q')=\frac{\mathcal{P}(q'\mid q,\zeta)\cdot |\pi_{q'}|_{\mathrm{cached}}}{\max\!\left(t_{\mathrm{reply}}+\hat{g}-t_{\mathrm{now}},\ \tau_{\min}\right)}
\]
where $|\pi_{q'}|_{\mathrm{cached}}$ is the number of cached tokens of $\pi_{q'}$, and $\tau_{\min}$ is the floor that prevents an imminent call from receiving an unbounded $\mathcal{D}$. \sysname admits predictions into \emph{PLANNED} in descending order of $\mathcal{D}$ until they cover a fixed fraction of the device tier. Within \emph{PLANNED}, each radix-cache node accumulates the $\mathcal{D}$ of every planned prefix that covers it, so a shared head outranks a private one.

\paragraph{Prefetch}
When part of $\pi_{q'}$ has been demoted to the host, the planner prefetches it back to the device ahead of the estimated arrival time. The prefetch release deadline is: 
\[
t_{\mathrm{release}}=t_{\mathrm{reply}}+\hat{g}-\frac{|\pi_{q'}|_{\mathrm{host}}}{B_{\mathrm{h2d}}}-\frac{|\pi_{q'}|_{\mathrm{uncached}}}{B_{\mathrm{prefill}}}
\]
where $B_{\mathrm{h2d}}$ and $B_{\mathrm{prefill}}$ are the measured host-to-device copy rate and prefill rate, $|\pi_{q'}|_{\mathrm{host}}$ is the number of tokens of $\pi_{q'}$ cached only on the host, and $|\pi_{q'}|_{\mathrm{uncached}}$ is the number of uncached tokens that require prefill.

Among released predictions, the one with earliest deadline goes first, with ties broken by the expected saving $\mathcal{P}(q'\mid q,\zeta)\cdot|\pi_{q'}|_{\mathrm{cached}}$. A prefetch is released when the engine is idle or its deadline $t_{\mathrm{release}}$ approaches. It takes only free device memory, after reclaiming \emph{RETIRED} and \emph{TRANSIENT} entries. When free space is insufficient, prefetching pauses but keeps refreshing the recency of these cache entries so that they do not age out. 

\begin{figure*}[t]
    \centering
    \includegraphics[width=0.95\textwidth]{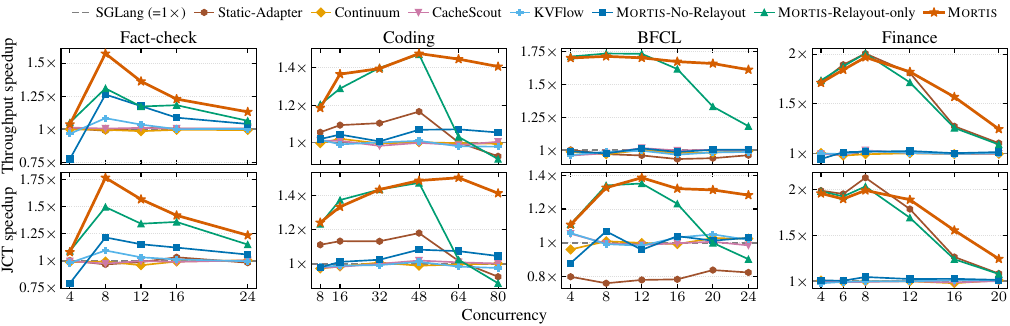}
    \vspace{-8pt}
    \caption{Throughput speedup (top) and average JCT speedup (bottom) over SGLang on each agent workload under varying concurrency.}
    \vspace{-8pt}
    \label{fig:e2e_speedup}
\end{figure*}

\section{Evaluation}
\label{sec:eval}
\subsection{Experimental Setup}
\para{Testbed and model}
We implement \sysname on top of SGLang v0.5.10.post1 and the Jac compiler~\cite{mars2023jaseci}, and conduct evaluations on one A40 GPU using Qwen3-32B-AWQ. We allocate 15\,GB of GPU memory and 25\,GB of CPU memory to the device and host KV-cache pools respectively. In \sysname, the retention fraction is set to 0.6 and $\tau_{\min}$ to 0.5\,s.

\para{Baselines}
We compare \sysname against five baselines, reimplemented on our SGLang fork:
\begin{enumerate*}[label=\textbf{(\roman*)}]
    \item \textbf{SGLang}~\cite{zheng2024sglang}, the vanilla SGLang engine with HiCache and default LRU eviction;
    \item \textbf{CacheScout}~\cite{zhang2026cachescout}, which learns call transitions online to prioritize and prefetch agent prefixes;
    \item \textbf{Continuum}~\cite{li2025continuum}, which pins agent KV-cache across tool gaps using adaptive TTLs;
    \item \textbf{KVFlow}~\cite{pan2025kvflow}, which schedules prefix eviction and prefetching via an Agent Step Graph; and
    \item \textbf{Static-Adapter}, which mirrors DSPy's adapter~\cite{dspy2026chatadapter} by hoisting invariant components to the front under a fixed, program-agnostic layout.
\end{enumerate*}

We also evaluate the following ablation variants of \sysname:
\begin{enumerate*}[label=(\roman*)]
    \item \textbf{\sysname-Relayout-only} enables adaptive prompt re-layout but replaces proactive KV-cache scheduling with LRU eviction;
    \item \textbf{\sysname-No-Relayout} enables proactive KV-cache scheduling but disables adaptive prompt re-layout; and
    \item \textbf{\sysname-No-Prefetch} disables only KV-cache prefetching.
\end{enumerate*}
Additionally, we run baseline serving systems on top of \sysname's re-layout to isolate scheduling policy under the same prompt layout.

\para{Workloads}
We evaluate \sysname on four representative agent workloads, ported to Jac byLLM: a HoVer-style~\cite{jiang2020hover} fact-check agent, a HumanEval~\cite{chen2021evaluating} coding agent, a BFCL~\cite{patil2025bfcl} tool-use agent, and a FinanceBench~\cite{islam2023financebench} financial-analysis agent. We further evaluate a mixed workload comprising all four agent types to assess \sysname's end-to-end performance under concurrent, heterogeneous demand. We report the average job completion time (JCT) and throughput (prompt tokens served per minute).

\subsection{End-to-End Performance of \sysname}
\para{Performance on different workload types}
As shown in Figure~\ref{fig:e2e_speedup}, \sysname achieves a 1.08--1.99$\times$ JCT speedup and increases throughput by 1.04--1.97$\times$ compared with SGLang, with geometric-mean gains of 1.45$\times$ in JCT and 1.49$\times$ in throughput. Across these agents, serving systems such as KVFlow show no significant improvement, because they ignore intra-prompt heterogeneity and the identified prefixes they operate on are short. 

Under low concurrency, adaptive prompt re-layout accounts for nearly all of the improvement, because memory is sufficient. As load increases, however, the KV-cache becomes contested. Without knowing what to keep and what to give up, other serving systems fail to schedule it efficiently. 

For the financial-analysis agent, KV-cache scheduling adds little, because this agent is a simple pipeline that digests large blobs of documents. Under low pressure, adaptive prompt re-layout accounts for nearly all of the improvement by exposing larger prefix reuse. \sysname's proactive KV-cache scheduling only takes effect under the highest load, because it can proactively evict stale KV-cache entries. 

\para{Performance on mixed workload}
We evaluate \sysname under varying pressure on the mixed workload and report metrics over each system's steady-state window. We vary the number of lanes (concurrent workflows) per agent, keeping it equal across all four agents.

\begin{figure}[tb]
    \centering
    \captionsetup[subfigure]{font=footnotesize,skip=1pt,margin=-6pt}
    \includegraphics[width=\columnwidth]{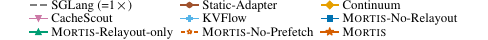}\\[1pt]
    \begin{subfigure}[t]{0.333\columnwidth}
        \centering
        \includegraphics[width=\linewidth]{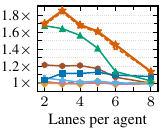}
        \caption{Throughput speedup}
        \label{fig:lane_sweep_tp}
    \end{subfigure}%
    \begin{subfigure}[t]{0.333\columnwidth}
        \centering
        \includegraphics[width=\linewidth]{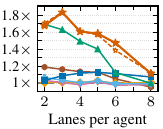}
        \caption{JCT speedup}
        \label{fig:lane_sweep_jct}
    \end{subfigure}%
    \begin{subfigure}[t]{0.333\columnwidth}
        \centering
        \includegraphics[width=\linewidth]{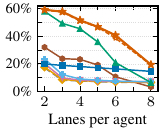}
        \caption{Cache hit rate}
        \label{fig:lane_sweep_hit}
    \end{subfigure}
    \vspace{-6pt}
    \caption{Throughput and JCT speedup over SGLang and cache hit rate under varying lanes per agent on the mixed workload. JCT speedup is the geometric mean of the four agents' speedups.}
    \label{fig:lane_sweep}
    \vspace{-8pt}
\end{figure}

As shown in Figure~\ref{fig:lane_sweep}, \sysname achieves up to 1.86$\times$ throughput speedup and 1.83$\times$ JCT speedup over SGLang at 3 lanes per agent. Under low pressure, adaptive prompt re-layout alone accounts for nearly all of the gain, since device memory is still large enough to retain the reusable KV-cache. As pressure grows, the cache becomes contested and the gain from re-layout alone shrinks, while proactive KV-cache scheduling retains a clear advantage by protecting the reusable heads that LRU would evict. Under the heaviest pressure, configurations converge toward the baseline, as the engine is fully saturated and no free memory is left for cache management to exploit. Across the six lane counts, \sysname achieves geometric-mean speedups of 1.56$\times$ in throughput and 1.53$\times$ in JCT.

In the mixed workload, agents have heterogeneous demands for KV-cache memory, which reduces the chance that they contend for memory at the same time. By exploiting intra-prompt heterogeneity, \sysname can promptly yield stale KV-cache to other requests that need memory, which further amplifies the benefit of \sysname.

\para{Task quality} The performance gains of \sysname do not come at the cost of task quality. As shown in Table~\ref{tab:layout_accuracy}, across three different layouts produced by the original byLLM, Static-Adapter, and \sysname, reordering components does not harm task quality. Moreover, the three layouts differ by at most 3\% in LLM calls per task and 6\% in output tokens. \sysname's speedup is not the result of less work either.

\begin{table}[tb]
    \centering
    \small
    \caption{Task accuracy under three prompt layouts that appear in our evaluation.}
    \label{tab:layout_accuracy}
    \resizebox{\linewidth}{!}{%
        \begin{tabular}{lcccc}
            \toprule
            Prompt layout & Fact-check & Coding & Financial-analysis & BFCL \\
            \midrule
            Original byLLM & 16.3\% & 55.3\% & 34.9\% & 14.1\% \\
            \sysname & 21.9\% & 54.8\% & 37.2\% & 18.7\% \\
            Static-Adapter  & 16.3\% & 52.2\% & 37.8\% & 16.0\% \\
            \bottomrule
        \end{tabular}%
    }
    \vspace{-8pt}
\end{table}

\begin{figure}[tb]
    \centering
    \includegraphics[width=\columnwidth]{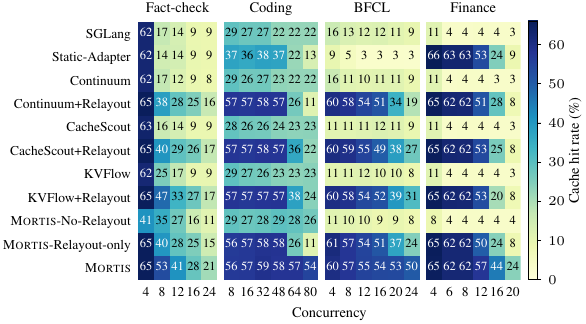}
    \caption{Cache hit rate (\%) of prompt tokens under varying concurrency}
    \vspace{-8pt}
    \label{fig:hit_rate}
\end{figure}

\subsection{Cache Efficiency of \sysname}
Figure~\ref{fig:hit_rate} shows the cache hit rate of prompt tokens at the same concurrency levels as Figure~\ref{fig:e2e_speedup}. On Coding and BFCL, adaptive prompt re-layout alone brings the hit rate at low concurrency to that of \sysname: it places the stable and shared bindings at the front, where they form a reusable head that later calls hit. As concurrency grows, the hit rate of \sysname-Relayout-only drops quickly, because LRU eviction does not distinguish reusable heads from volatile tails. With proactive KV-cache scheduling, \sysname keeps a higher hit rate than \sysname-Relayout-only as concurrency grows, and on Coding its hit rate stays nearly flat. On the Fact-check agent, \sysname's hit rate exceeds that of \sysname-Relayout-only from the lowest concurrency. Continuum and CacheScout stay close to SGLang at every concurrency. \sysname-No-Relayout stays below \sysname, since under the original prompt order the reusable head is short. The mixed workload shows a similar pattern (Figure~\ref{fig:lane_sweep}).

\subsection{Ablation Study}
As the geometric-mean speedups of JCT and throughput over SGLang on the four workloads in Table~\ref{tab:complementary} show, \emph{adaptive prompt re-layout} and \emph{proactive KV-cache scheduling} are superadditive: when both are enabled, the speedup is higher than the product of the two individual speedups. The same effect appears on the mixed workload (Figure~\ref{fig:lane_sweep_tp}), where adding proactive KV-cache scheduling on top of adaptive prompt re-layout substantially improves performance.

The gain of \sysname's KV-cache scheduling builds on re-layout, because re-laying out the prompt order exposes a reusable prefix and a suffix that can be abandoned. \sysname's KV-cache scheduling substantially improves performance on top of re-layout, but scheduling alone (\sysname-No-Relayout) barely helps (Table~\ref{tab:complementary}). The same trend appears in Figures~\ref{fig:lane_sweep} and \ref{fig:e2e_speedup}.

Prefetching contributes little on this testbed. As shown in Figure~\ref{fig:lane_sweep_tp}, disabling it (\sysname-No-Prefetch) lowers throughput by at most 2.6\% across the six lane counts. The gain of proactive KV-cache scheduling mainly comes from retention, which keeps the reusable heads resident.

\begin{table}[t]
    \centering
    \caption{Geometric-mean speedup over SGLang across all concurrency levels of the four agent workloads, with \emph{Adaptive Prompt Re-layout} and \emph{Proactive KV-cache Scheduling} on or off.}
    \label{tab:complementary}
    \small
    \setlength{\tabcolsep}{4pt}
    \begin{subtable}[t]{0.49\columnwidth}
        \centering
        \caption{JCT speedup}
        \begin{tabular}{lcc}
        \toprule
         & \multicolumn{2}{c}{Scheduling} \\
        \cmidrule(lr){2-3}
        Re-layout & Off & On \\
        \midrule
        Off & 1.00$\times$ & 1.03$\times$ \\
        On  & 1.30$\times$ & 1.45$\times$ \\
        \bottomrule
        \end{tabular}
    \end{subtable}
    \hfill
    \begin{subtable}[t]{0.49\columnwidth}
        \centering
        \caption{Throughput speedup}
        \begin{tabular}{lcc}
        \toprule
         & \multicolumn{2}{c}{Scheduling} \\
        \cmidrule(lr){2-3}
        Re-layout & Off & On \\
        \midrule
        Off & 1.00$\times$ & 1.02$\times$ \\
        On  & 1.36$\times$ & 1.49$\times$ \\
        \bottomrule
        \end{tabular}
    \end{subtable}
    \vspace{-8pt}
\end{table}

\subsection{Comparison under Identical Layouts}
\begin{figure}[t]
    \centering
    \captionsetup[subfigure]{font=footnotesize,skip=1pt}
    \includegraphics[width=\columnwidth]{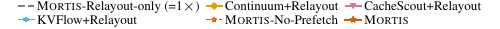}\\[1pt]
    \begin{subfigure}[t]{\columnwidth}
        \centering
        \includegraphics[width=\linewidth]{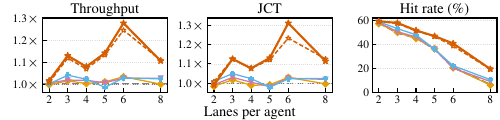}
        \caption{Throughput, JCT speedup and hit rate on the mixed workload}
        \label{fig:mixed_baseline_relayout}
    \end{subfigure}\\[1pt]
    \begin{subfigure}[t]{\columnwidth}
        \centering
        \includegraphics[width=\linewidth]{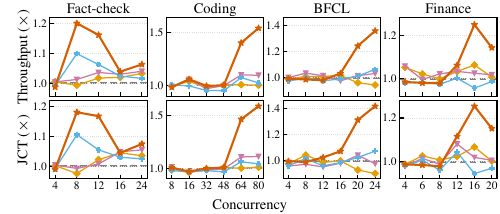}
        \caption{Throughput and JCT speedup on each agent workload}
        \label{fig:speedup_baseline_relayout}
    \end{subfigure}
    \vspace{-6pt}
    \caption{Baseline systems served with \sysname's re-laid prompts. Speedups are measured against \sysname-Relayout-only.}
    \vspace{-8pt}
\end{figure}

Serving the baseline systems with re-laid prompts identical to \sysname's, we evaluate their speedup against \sysname-Relayout-only. As shown in Figure~\ref{fig:mixed_baseline_relayout}, these systems show marginal improvement over \sysname-Relayout-only, but are still beaten by \sysname. We further evaluate on the four agent workloads separately. Figure~\ref{fig:speedup_baseline_relayout} shows the same trend as in Figure~\ref{fig:mixed_baseline_relayout}. The largest gains are from CacheScout on Coding and KVFlow on Fact-check, both up to 1.11$\times$ in JCT, but \sysname still outperforms all of them.

This deficiency in baseline serving systems arises because they manage the KV-cache only at the granularity of a prefix and its suffix, which ignores intra-prompt heterogeneity. They cannot identify precisely which KV-cache entries will be reused, leading to inefficient KV-cache scheduling.

\subsection{Sensitivity Study}
\begin{figure}[t]
    \centering
    \captionsetup[subfigure]{font=footnotesize,skip=1pt,margin=-6pt}
    \includegraphics[width=\columnwidth]{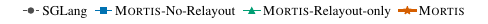}\\[0pt]
    \begin{subfigure}[t]{0.5\columnwidth}
        \centering
        \includegraphics[width=\linewidth]{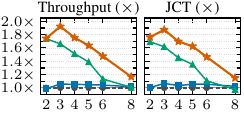}
        \caption{A40 TP=4}
    \end{subfigure}%
    \begin{subfigure}[t]{0.5\columnwidth}
        \centering
        \includegraphics[width=\linewidth]{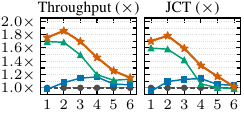}
        \caption{RTX 3090}
    \end{subfigure}
    \vspace{-9pt}
    \caption{Throughput and JCT speedup over SGLang on the mixed workload under varying lanes per agent.}
    \label{fig:sensitivity_study}
    \vspace{-12pt}
\end{figure}

We run the mixed workload on four A40 GPUs with TP=4 and the same KV-cache pool size configurations. TP=4 increases the prefill speed by 1.74$\times$ (from 1292.4 to 2251.8 tokens/s), which lowers the cost of a cache miss. As shown in Figure~\ref{fig:sensitivity_study}, \sysname maintains its advantage even under cheaper cache misses. It achieves geometric-mean speedups of 1.60$\times$ in throughput and 1.58$\times$ in JCT (up to 1.92$\times$ and 1.87$\times$). Additionally, we evaluate on a smaller GPU with a smaller model. On one RTX 3090 with Qwen3-14B-AWQ, with 5.2\,GB of GPU memory and 10\,GB of CPU memory for the KV-cache pools, \sysname keeps the same pattern of improvement (Figure~\ref{fig:sensitivity_study}): re-layout exposes additional prefix cache reuse, and KV-cache scheduling takes effect once the cache becomes contested and thrashes.

\section{Related Work}
\label{sec:related_works}
\para{Application-level information for serving} Parrot~\cite{lin2024parrot} represents intermediate results as semantic variables and uses their producer--consumer relationships for scheduling and prefix-aware placement. Teola~\cite{tan2025teola} exposes operators finer than a call, removes unnecessary dependencies, computes available prefixes while upstream operators produce the remaining inputs, and forwards partial outputs downstream. 

\para{Prompt construction for prefix reuse} DSPy's adapters and NOOA fix a component order by convention or by developer labels~\cite{dspy2026chatadapter,furgale2026nooa}, and the SGLang frontend moves constant components forward at compile time~\cite{zheng2024sglang}. For relational analytics, \citet{liu2025optimizing} order rows and the fields within them jointly, grouping rows that share field values and using functional dependencies to expose further sharing. Reuse can also go beyond matching prefixes: Prompt Cache~\cite{gim2024promptcache} precomputes prompt modules at prescribed positions, and CacheBlend~\cite{yao2025cacheblend} recomputes the tokens whose attention crosses chunk boundaries when combining cached chunks.
%  \sysname adjusts the framework's lowering process according to analysis of the agent program itself to increase prefix sharing.

\para{KV-cache scheduling across execution gaps} Serving systems also steer the KV-cache according to tool-call gaps in ReAct agents. InferCept~\cite{abhyankar2024infercept} weighs the cost of keeping, swapping, and recomputing KV-cache per interruption to make scheduling decisions; Continuum~\cite{li2025continuum} holds the idle KV-cache for a time-to-live derived from observed inter-invocation delays; TokenCake~\cite{bian2026tokencake} estimates the tool call's duration, offloads idle KV to the host, and restores it before the expected return. 
% They focus on how to handle the idle KV-cache to maximize management gains.
\section{Conclusion}
\label{sec:conclusion}
% We present \sysname, a co-design of the compiler and serving runtime that drives more efficient KV-cache scheduling by exploiting intra-prompt heterogeneity. It uses runtime-assisted static analysis and steers the KV-cache through prompt re-layout and proactive scheduling. Our evaluation shows that \sysname achieves up to 1.99$\times$ JCT speedup and 1.97$\times$ higher throughput over SGLang. \sysname highlights the importance of the structure inside prompts for serving.
\sysname exploits intra-prompt heterogeneity via runtime-assisted static analysis, steering KV-cache management through adaptive prompt re-layout and proactive scheduling. Evaluations show \sysname achieves up to 1.99$\times$ JCT speedup and 1.97$\times$ higher throughput over SGLang.

\bibliographystyle{mlsys2025}
\bibliography{references}

\section*{AI Usage Disclosure}
In this paper, we use Claude Code for prototyping and methodology investigation and organization. During system development, we use AI to help with algorithm design and implementation. In the paper-writing process, we equip Claude Code with grammar-checking skills to clean up glitches in writing.

\end{document}